\documentclass[aps,prl,twocolumn,superscriptaddress]{revtex4-1}
\usepackage{graphicx}
\usepackage{color}
\usepackage{amsfonts}
\usepackage{ulem}

\definecolor{brown}{RGB}{200,100,0}

\def\WS2{WS$_2$}
\def\MoS2{MoS$_2$}{

\begin{document}
\title{Accessing the spectral function in a current-carrying device}
\author{ Davide Curcio}
\author{Alfred J. H. Jones}
 \affiliation{Department of Physics and Astronomy, Interdisciplinary Nanoscience Center, Aarhus University,
8000 Aarhus C, Denmark}
\author{Ryan Muzzio}
 \affiliation{Department of Physics, Carnegie Mellon University, Pittsburgh PA 15213, USA}
\author{Klara Volckaert}
\author{Deepnarayan Biswas}
 \affiliation{Department of Physics and Astronomy, Interdisciplinary Nanoscience Center, Aarhus University,
8000 Aarhus C, Denmark}
\author{Charlotte E. Sanders}
\affiliation{Central Laser Facility, STFC Rutherford Appleton Laboratory, Harwell, United Kingdom}
\author{Pavel Dudin}
\author{Cephise Cacho}
\affiliation{Diamond Light Source, Division of Science, Didcot, United Kingdom}
\author{Simranjeet Singh}
 \affiliation{Department of Physics, Carnegie Mellon University, Pittsburgh PA 15213, USA}
\author{Kenji Watanabe}
\author{Takashi Taniguchi}
\affiliation{National Institute for Materials Science, Tsukuba 305-0044, Japan}
\author{Jill A. Miwa}
\affiliation{Department of Physics and Astronomy, Interdisciplinary Nanoscience Center, Aarhus University,
8000 Aarhus C, Denmark}
\author{Jyoti Katoch}
 \affiliation{Department of Physics, Carnegie Mellon University, Pittsburgh PA 15213, USA}
\author{ S\o ren Ulstrup}
\author{ Philip Hofmann}
\affiliation{Department of Physics and Astronomy, Interdisciplinary Nanoscience Center, Aarhus University,
8000 Aarhus C, Denmark}
\begin{abstract}
The presence of an electrical transport current in a material is one of the simplest and most important realisations of non-equilibrium physics. The current density breaks the crystalline symmetry and can give rise to dramatic phenomena, such as sliding charge density waves \cite{Monceau:1976aa}, insulator-to-metal transitions \cite{Kumai:1999aa,Sow:2017ab} or gap openings in topologically protected states \cite{Balram:2019aa}. Almost nothing is known about how a current influences the electron spectral function, which characterizes most of the solid's electronic, optical and chemical properties. Here we show that angle-resolved photoemission spectroscopy with a nano-scale light spot (nanoARPES) provides not only a wealth of information on local equilibrium properties, but also opens the possibility to access the local non-equilibrium spectral function in the presence of a transport current. Unifying spectroscopic and transport measurements in this way allows non-invasive local measurements of 
the composition, structure, many-body effects and carrier mobility in the presence of high current densities.
\end{abstract}
\maketitle 

The spectral function of a solid encodes its electronic properties, including many-body effects, and is therefore of key interest for understanding a vast range of physical and chemical properties. The dominant experimental technique for determining the spectral function is angle-resolved photoemission spectroscopy (ARPES) \cite{Damascelli:2003aa} but so far few attempts have been made to carry out such measurements in the presence of a transport current, mainly because of the significant voltage drop over the area of the UV light spot used for photoexcitation, and the detrimental effect this has on the energy resolution (except in the case of superconductors \cite{Kaminski:2016aa,Naamneh:2016aa}). The use of a nanoscale light spot circumvents this voltage drop problem and thus permits the determination of the local spectral function in the presence of a current. 

\begin{figure*}
\begin{center}
\includegraphics[width=0.75\textwidth]{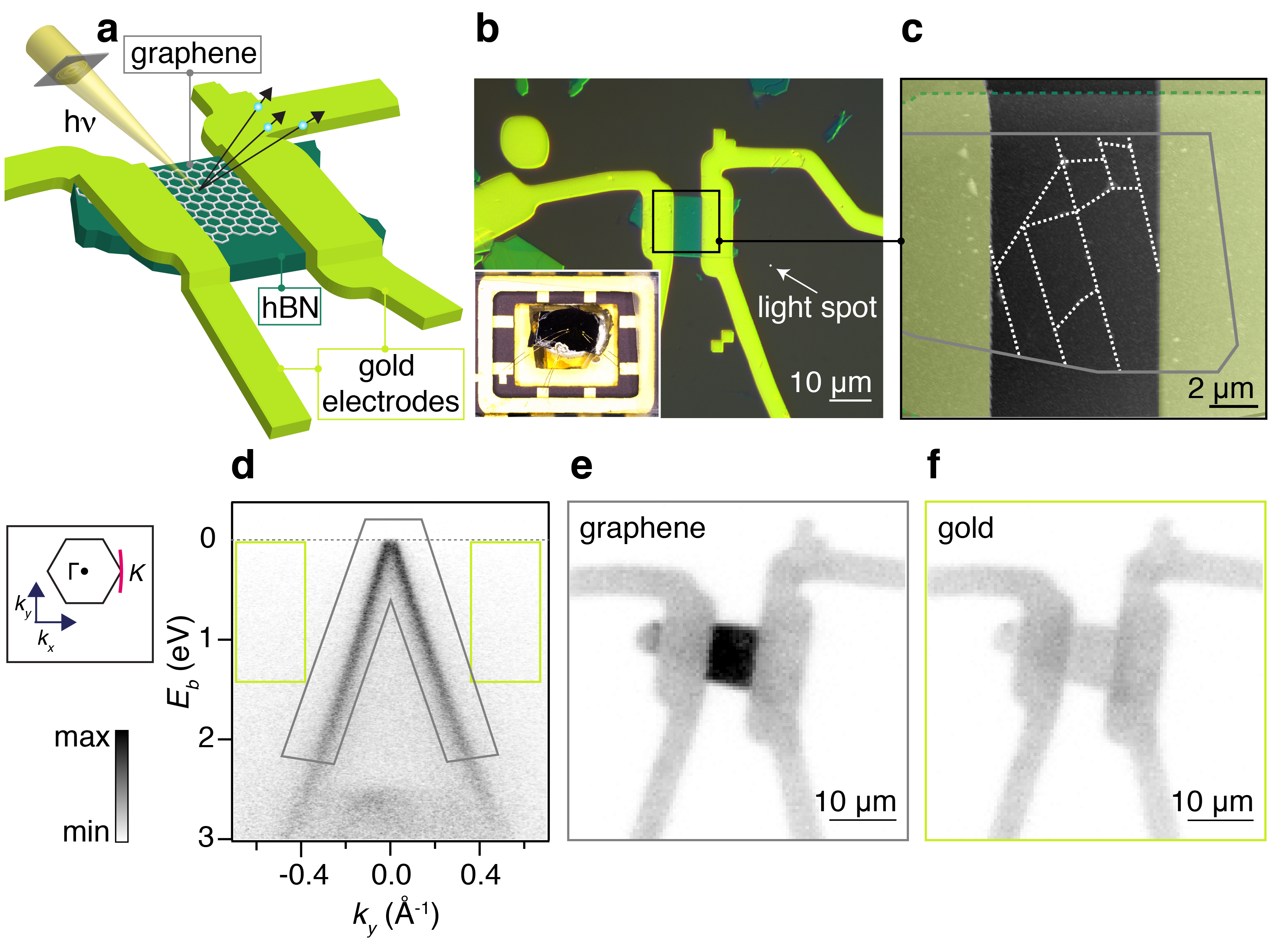}
\caption{\textbf{Characterization of a graphene / hBN device.} \textbf{a} Sketch of the device with the UV beam focused onto it using a zone plate. \textbf{b} Optical image of the device. The white dot gives the size of the light spot used (diameter of 500~nm) with respect to the device. The inset shows the entire device as it is mounted on a chip carrier. \textbf{c} Atomic force microscopy image of the device together with an outline of the hBN flake (dashed green line) and graphene (grey line). The white dashed lines follow rows of protrusions identified on the surface by AFM (see Appendix). The size of the graphene flake is determined by nanoARPES, see panel e.  \textbf{d} Spectrum taken from the middle of the device (photoemission intensity as a function of binding energy $E_b$ and $k_y$), showing the Dirac cone of graphene and the valence band maximum of hBN at $E_b \approx$2.7~eV.  The sketch to the left shows the Brillouin zone together with the scan direction (red line). \textbf{e} Integrated photoemission intensity in the grey window around the Dirac cone in panel d, marking the location of graphene.  \textbf{f} Photoemission intensity from the green area in panel d, emphasizing the Au contacts.}
\label{fig:1}
\end{center}
\end{figure*}

We demonstrate access to detailed local equilibrium and non-equilibrium properties using a device consisting of exfoliated single layer graphene, placed on a 28~nm thick hexagonal boron nitride (hBN) flake on a SiO$_2$ substrate (see Figs. \ref{fig:1}a and b). The graphene flake is electrically contacted by two Au electrodes, permitting the application of a lateral source-drain voltage. Fig. \ref{fig:1}c shows an atomic force microscopy (AFM) image of the device. The area of the graphene flake as determined by nanoARPES (see below) is outlined. The white dashed lines within the area of the graphene flake track rows of protrusions on the surface (see Appendix for an AFM image without these lines, showing the protrusions more clearly). The lines serve to distinguish between  possibly different regions within the graphene. Fig. \ref{fig:1}d shows the energy- and $k$-resolved photoemission intensity taken near the centre of the device, revealing the  Dirac cone of graphene, as well as the valence band maximum of hBN  \cite{Koch:2018aa}. By scanning the light spot across the sample, the intensity of characteristic spectroscopic features can be mapped. Fig. \ref{fig:1}{e} shows the intensity of the Dirac cone (integrated within the grey outlined region of Fig. \ref{fig:1}d), revealing the location of the graphene flake within the device. The resulting shape has also been used for the outline of the flake in Fig. \ref{fig:1}c.  Fig. \ref{fig:1}f shows the intensity within the green marked regions of the spectrum in Fig. \ref{fig:1}d. This emphasizes the Au contacts more strongly due to the absence of graphene spectral features. 

\begin{figure}
\begin{center}
\includegraphics{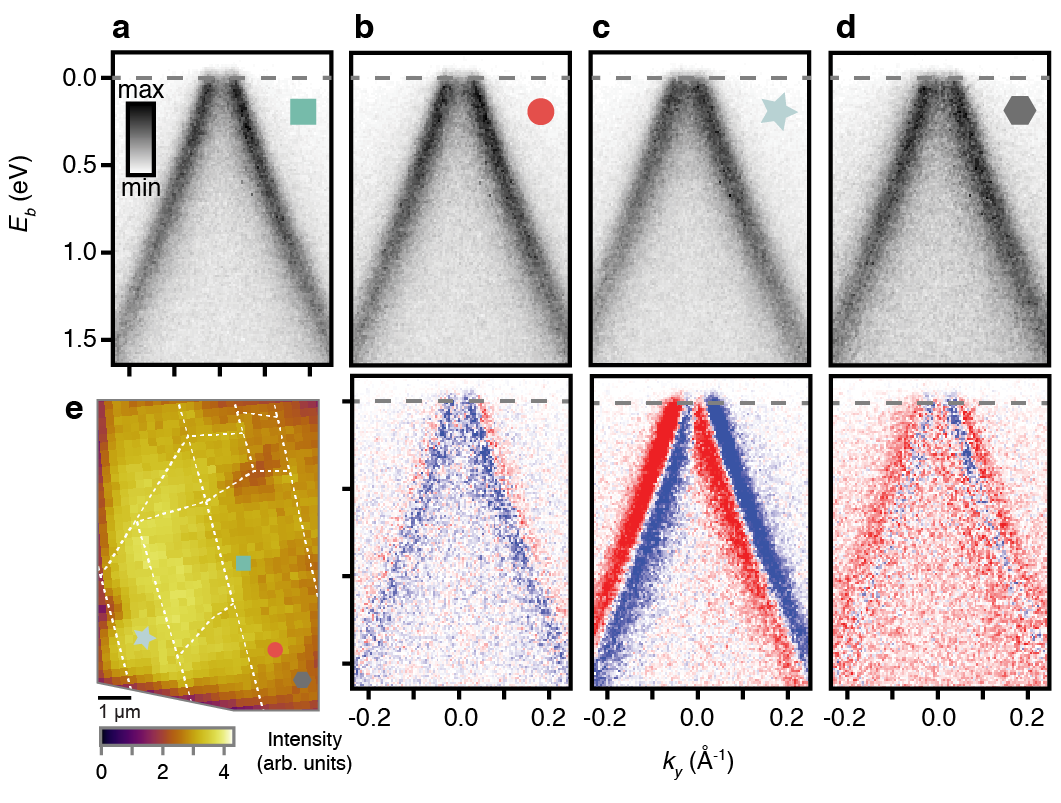}
\caption{\textbf{Graphene spectra taken in different locations of the device}, as given by the corresponding marker in panel e.  \textbf{a} Reference spectrum. \textbf{b}-\textbf{d} Spectra taken at different positions along with the difference to the reference spectrum below (blue negative, red positive). From the difference plots, it can be concluded that the spectrum in panel b is shifted upwards in energy, the spectrum in panel c is shifted towards negative $k_y$ values and the spectrum in panel d is broader than the reference spectrum,  as seen by the intensity decrease in the middle of the branches and the increase in the tails.  \textbf{e} Intensity of the Dirac cone across the device, obtained by the fitting procedure also used for Fig. \ref{fig:3}, with the locations for the spectra in panels a-d given by coloured markers.
}
\label{fig:2}
\end{center}
\end{figure}

The observed spectral features show some variation across the device. This is best seen by using an arbitrary spectrum (in Fig. \ref{fig:2}a) as a reference and comparing this to spectra taken at other locations. Fig. \ref{fig:2}b-d give such spectra, along with the difference spectrum relative to the reference (for the corresponding real-space locations on the device, see Fig. \ref{fig:2}e). Evidently, the spectrum in Fig. \ref{fig:2}{b} is  shifted in energy, the spectrum in Fig. \ref{fig:2}{c} is  shifted in $k_y$ and the spectrum in Fig. \ref{fig:2}{d} shows slightly broader lines, corresponding to a reduced quasiparticle lifetime. 

For a more detailed analysis we fit the spectra across the sample using a simple model spectral function $\mathcal{A}(E_b,k_y)$ of the form 
 \begin{equation}
  \mathcal{A}(E_b,k_y)=\frac{\pi^{-1}|\Sigma''|}{[E_b-\epsilon(k_y)]^{2}+\Sigma''^{2}}, 
 \label{equ:p7}
 \end{equation}
where  $\epsilon(k_y)$ is the Dirac cone dispersion and  $\Sigma''$ is the imaginary part of the self energy that accounts for the broadening of the observed features. The experimental intensity is modelled as a product of the spectral function with a Fermi-Dirac distribution and a branch-dependent intensity, convoluted with an experimental resolution function. It proves advantageous to simultaneously fit the complete 2D intensity image rather than merely one-dimensional cuts at certain energy or $k_y$-values \cite{Nechaev:2009aa} (see Appendix  for further details on the analysis procedure).

\begin{figure}
\begin{center}
\includegraphics{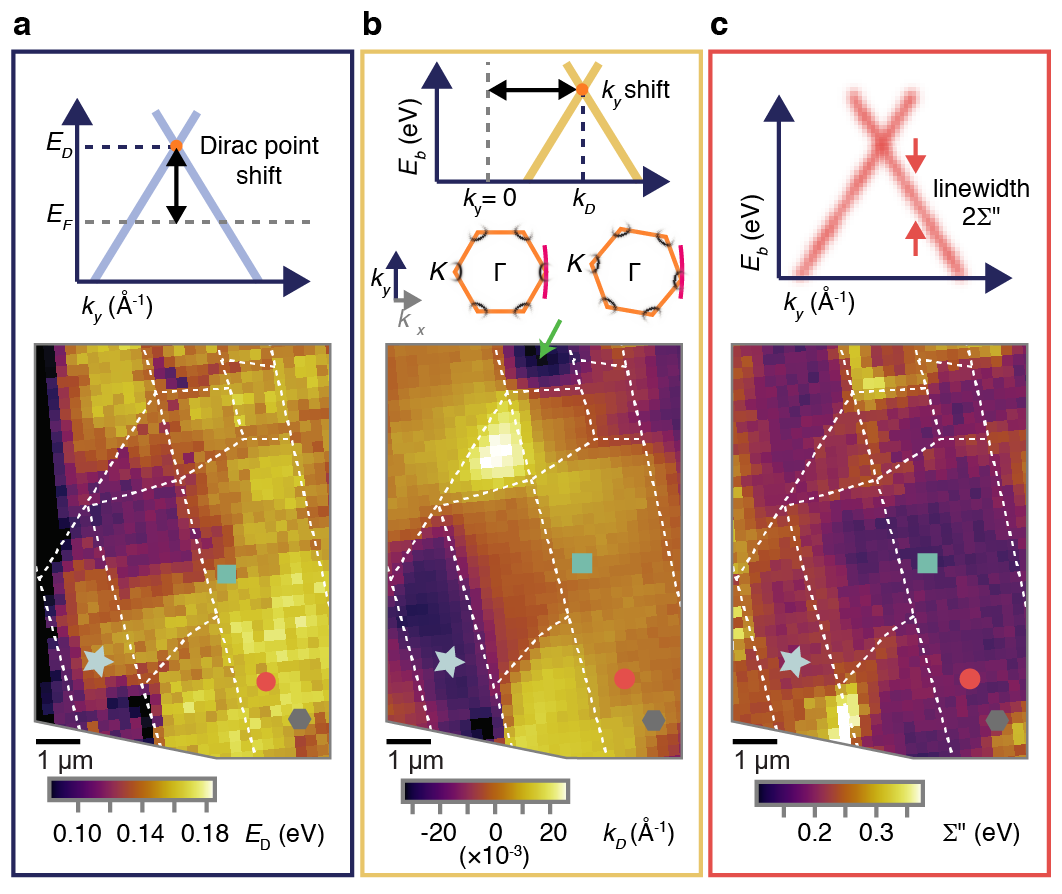}
\caption{\textbf{Detailed characterization of the device properties}  obtained by scanning the light spot across the device and fitting the spectra measured at every point. The upper parts of the sub-figures schematically show the property encoded in the measured parameter. \textbf{a} Energy of the Dirac point $E_D$, corresponding to the local variation in doping across the device. The markers correspond to the locations for the spectra in Fig. \ref{fig:2}. \textbf{b} Position of the Dirac cone in $k_y$. A shift is primarily assigned to azimuthal rotations between domains on the sample. The inset demonstrates how a domain rotation from a reference direction leads to a shift of the Dirac cone dispersion with respect to the fixed scan range in $k-$space (red line). Note that the rotation between domains in the sketch (8$^{\circ}$) is much larger than observed here in order to illustrate the effect. \textbf{c} Imaginary part of the self energy $\Sigma''$ measuring the width of the spectra or, equivalently, the inverse lifetime of the state.  }
\label{fig:3}
\end{center}
\end{figure}

Fig. \ref{fig:3} shows the key parameters of this fit mapped across the device in equilibrium. The Dirac point energy $E_D$, corresponding to the local doping, varies across the surface of the graphene flake, lying mostly between $\approx0.12$ and $\approx0.18$~eV above the Fermi level (see Fig. \ref{fig:3}{a}). This corresponds to a hole density between $\approx7 \times10^{11}$ and $\approx17 \times10^{11}$cm$^{-2}$. We ascribe the hole doping to the presence of residual water on the surface \cite{Wang:2010ad}.

Fig. \ref{fig:3}{b} shows the location-dependent position of the Dirac cone in $k_y$. This map shows distinct and well-defined regions that coincide remarkably well with the dashed lines on the AFM image of Fig. \ref{fig:1}c, which are also overlaid on the data in Fig. \ref{fig:3} . We assign the displacement in $k_y$ for different sample regions to the presence of domains that are rotated against each other by a small angle.  To see this, consider the two orange Brillouin zones in Fig. \ref{fig:3}{b} that have slightly different azimuthal orientations with respect to one another. This leads to an  apparent $k$-shift of the Dirac cone in the window of observation (marked by a red line). 
The interpretation in terms of rotational domains is supported by the intensity ratio between the two branches of the Dirac cone that can be used to determine the domain orientation relative to the  characteristic ``dark corridor'' in the photoemission intensity from graphene \cite{Shirley:1995aa,Mucha-Kruczynski:2008aa,Lizzit:2010aa,Gierz:2011ab}, see Appendix. The orientational disorder between the domains is very small, with the full range of Fig.  \ref{fig:3}{b} corresponding to $\approx2^{\circ}$.  We stress that such a small variation of the local azimuthal orientation is still very important for, e.g., the properties of twisted bilayer graphene \cite{Kerelsky:2019aa}. Based on the interpretation of the $k_y$ shift in terms of rotational domains, the lines of protrusions in the AFM image are likely to be caused by wrinkles or folds in the graphene sheet that result in slightly different rotational orientations on either side of theses boundaries.

The linewidth as measured by $\Sigma''$ in Fig.  \ref{fig:3}{c} shows large uniform regions of lower $\Sigma''$ (e.g., around the  square marker) but also smaller areas with significantly higher values,  such as the region with the large $k_y$ shift around the star-shaped marker, or the lower right corner of the flake near the hexagonal marker. We also observe an increase of $\Sigma''$ at the boundaries between some of the rotational domains. 
This is clearly seen for the  domain near the upper device edge (see green arrow in Fig.  \ref{fig:3}{b)} which is surrounded by a border of high $\Sigma''$ in Fig.  \ref{fig:3}{c}. Such an increase of $\Sigma''$ close to a domain boundary is, however, not necessarily due to a true lifetime decrease but can rather be caused by the simultaneous sampling over two domains, such that the sum of two slightly displaced Dirac cones is measured, leading to a spuriously large linewidth. Note that the mapping of $\Sigma''$ presented here is a proof of principle, illustrating the possibility to map the effect of many-body interactions. Nevertheless, while  $\Sigma''$ as presented here is an average over the broad energy region of the fit  (500~meV) and thus hides subtle details of, e.g., renormalization due to the electron-phonon interaction, it is still a meaningful measure of the -- largely energy independent -- lifetime reduction by electron-defect scattering \cite{Hofmann:2009ab}. Improved and more efficient optics for focusing the UV beam promise to give orders of magnitude higher photon flux \cite{Rotenberg:2014aa}, such that nanoARPES will eventually be able to map the same subtle details in the self energy that are now routinely observed by high resolution ARPES \cite{Damascelli:2003aa,Bostwick:2007aa}. It should then be possible to measure effects such as the dependence of electron-electron or electron-phonon coupling strength on the distance from a defect or a grain boundary. 
 
 \begin{figure*}
\begin{center}
\includegraphics[width=0.9\textwidth]{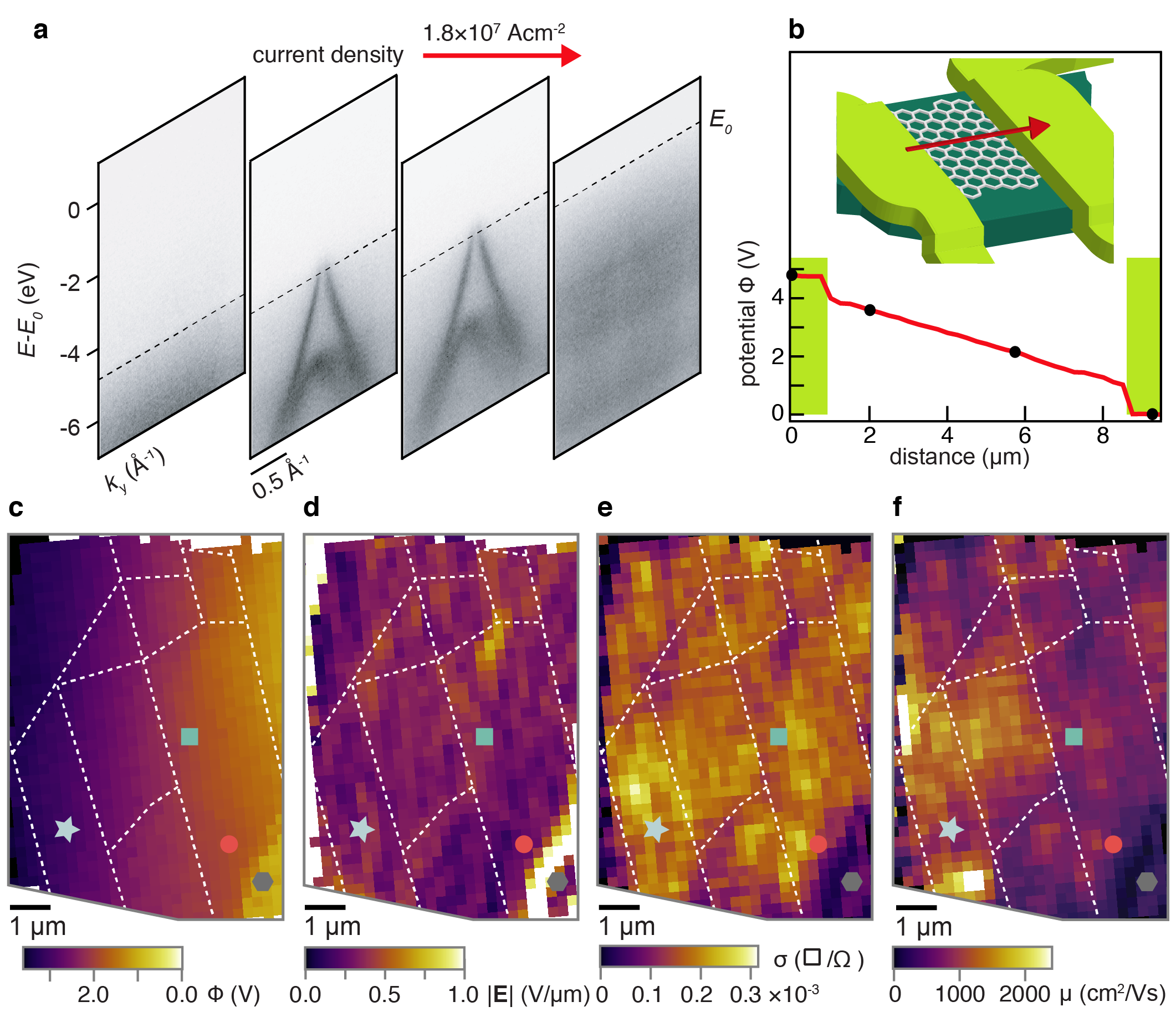}
\caption{\textbf{Analysis of current-carrying device.} \textbf{a} Series of spectra along a line connecting the Au electrodes (see inset in panel b) for a device current of 0.5~mA. The Fermi level on the right Au electrode is used as a zero for the energy scale.  \textbf{b} Sketch of the device along with the potential $\phi$ along the red line, determined from the local Fermi energy. The black markers give the location of the spectra shown in part a. $\phi$ shows a linear change over the graphene region and steep drops at the interfaces to the Au, corresponding to the contact resistances. \textbf{c} Map of the potential $\phi$ across the device for a current of 0.5~mA, also determined from the local Fermi energy. The energy zero is the Fermi energy on the right hand side. The markers correspond to the locations for the spectra in Fig. \ref{fig:2}.  \textbf{d} Magnitude of the electric field $|\mathbf{E}|=|\nabla \phi |$ across the device.  \textbf{e} Local conductivity $\sigma$ calculated from $\phi$. \textbf{f} Local mobility $\mu$, calculated by combining $\sigma$ with the hole density derived from the Dirac point energy $E_D$ in Fig. \ref{fig:3}a.   
}
\label{fig:4}
\end{center}
\end{figure*}
 
We proceed with the demonstration that, due to the small light spot in nanoARPES, it remains possible to measure the spectral function in the presence of a steady state current. Fig. \ref{fig:4}{a} shows a series of spectra taken along a line connecting both electrodes (see sketch in the inset of Fig. \ref{fig:4}{b}) for an applied current of 0.5~mA. The first and last spectrum are taken on the Au electrodes. Due to the electrodes being polycrystalline, no distinct bands are visible and the only features are the Fermi edge and the Au $d$-band. The middle spectra show graphene features very similar to the equilibrium situation, i.e. unaffected by the current.  There is a large energy offset between the spectra, as easily seen by the shift in the local Fermi energy. This is a consequence of the current density and the accompanying voltage drop across the device. The local potential between the electrodes is shown in Fig. \ref{fig:4}b, obtained by tracking the Fermi level of all spectra along the red line connecting the electrodes in the sketch. We observe the expected linear potential change across the graphene flake along with sharper drops at the graphene/ Au boundaries. The voltage drop at the left boundary is $\approx0.75$~V, from which we obtain a contact resistance of $\approx14$~k$\Omega \mu$m, which is on the high end of the expected graphene/ metal contact resistance range \cite{Xia:2011aa}.

The non-equilibrium properties across the entire device in the presence of a current of 0.5~mA are analyzed in the same way as introduced in Fig. \ref{fig:3}.  Again, a fit is performed to every spectrum taken as the light spot is scanned across the surface. The position-dependent Fermi energy can be used as a measure of the local potential $\phi$ as shown in Fig. \ref{fig:4}{c}. Due to the strong overall change, the map appears relatively structureless apart from a bright area near the lower right corner (near the hexagonal marker). This part of the graphene flake appears to remain at the same potential as the Au electrode and is thus not connected to rest of the graphene, implying a major structural defect. The same region was found to stand out in terms of a high $\Sigma''$ (short lifetime) in Fig. \ref{fig:3}c, also indicating the presence of structural defects.   More details are revealed upon calculating $|\mathbf{E}|=|\nabla \phi |$, i.e. the magnitude of the local electric field (Fig. \ref{fig:4}{d}). As expected, the missing connection between the graphene flake and the triangular area near the hexagonal marker now shows up as a high local electric field. 

 Employing approaches developed in connection with scanning tunnelling potentiometry and related techniques \cite{Muralt:1986aa,Zhang:2016ag,Ella:2019aa}, we can determine the local conductivity $\sigma$ from the potential map in Fig. \ref{fig:4}{c}. The result of this procedure is shown in Fig. \ref{fig:4}{e}. There is a close correspondence between the  $|\mathbf{E}|$ and $\sigma$ maps, with a high electric field corresponding to a low conductivity and vice versa. Despite of the more complicated situation here, this is reminiscent of the homogeneous situation with $\mathbf{E}=\mathbf{j}/\sigma$. Remarkably, we also find a close relation  between the position of the dashed white lines and a high electric field/ a low conductivity. According to the discussion of the rotational domains on the sample, the white lines mark locations of wrinkles in the graphene sheet and a low conductivity is thus to be expected in their vicinity. The results from the transport experiment are therefore in excellent agreement with the static characterization of the device. 

 In contrast to scanning probe potentiometry, our technique not only provides the local potential but also the complete position-resolved spectroscopic information, so that we can combine the conductivity data in Fig. \ref{fig:4}{e} with the local hole density $p$, calculated from the Dirac point energy $E_D$ in Fig. \ref{fig:3}{a}, in order to calculate the local mobility  of the device $\mu(x,y) = \sigma(x,y)/p(x,y)e $. The result  is shown in Fig. \ref{fig:4}{f}. The influence of the local carrier density is clearly seen in the mobility map. The region around the square marker, for instance, shows a high conductivity but the mobility is higher still $\approx2$~$\mu$m to the left of the square marker because of the much lower doping in that region. 

Finally, we can relate the results of the non-invasive transport measurements in Fig. \ref{fig:4} to the lifetime measurements via $\Sigma''$ in Fig. \ref{fig:3}c. Mostly, the results show the expected correlation between transport properties on one hand and $\Sigma''$ on the other hand. The region to the left of the square marker, for instance, shows a low $\Sigma''$, consistent with low electron-defect scattering, along with a high mobility. The defective region around the hexagonal marker, on the other hand, shows a high $\Sigma''$ and a low mobility. 

In conclusion, we have shown that nanoARPES can give unprecedented information about the local electronic structure and many-body effects in a device, both in equilibrium and in the presence of a steady state current. While numerous techniques can be used to map properties such as the local potential, current density, carrier concentration, or even mobility \cite{Muralt:1986aa,Tokura:1996aa,Vasyukov:2013aa,Barreto:2013aa,Tetienne:2017aa,Ella:2019aa,Voigtlander:2018aa}, no other approach simultaneously provides all of these in addition to the sample's spectral function. Even mapping of devices in equilibrium along the lines shown here is likely to produce a wealth of information of local many-body effects and their relation to mesoscopic structures and defects in devices, such as domain boundaries, line defects and local doping profiles. The introduction of a steady current will then allow the relationship between such many-body effects and transport relevant properties to be unravelled on a local scale, as well as open the option to investigate electronic structure changes in the presence of current-induced phase transitions.

\section{Methods}
\textbf{Device fabrication} 
The graphene on hBN heterostructure was prepared using a polymer-based transfer technique.  hBN crystals were exfoliated onto a SiO$_2$/Si substrate. Flakes with a thickness between 20 and 30~nm  were identified using optical contrast under a microscope. Separately, single-layer graphene flakes were obtained from graphite exfoliated on different SiO$_2$/Si substrates.   Then polycarbonate on polydimethylsiloxane slides were used to pick up graphene from the substrate and transfer it onto selected hBN flakes of $\approx28$~nm thickness.
For the device fabrication, polymethyl methacrylate / methyl methacrylate bilayer polymers were spun coated onto the sample and standard electron-beam lithography was used to define the electrodes. An electron beam evaporator was then used to deposit Au (90~nm)/Cr (5~nm)  metal contacts to the graphene flake.

\textbf{nanoARPES} NanoARPES experiments were carried out at the I05 beamline of Diamond Light Source. The device was mounted on a chip carrier, as shown in the inset Fig. \ref{fig:1}b, and transferred to the sample manipulator of the nanoARPES end station. The sample temperature during the measurements was 80~K and the photon energy 60~eV. 
The standard scanning mode involved collecting photoemission spectra with a Scienta Omicron DA30 hemispherical analyzer by rastering the sample position with respect to the focused synchrotron beam in steps of 250 nm using SmarAct piezo stages. The actual device could easily be found using a coarse $x,y$-scan while monitoring the integrated photoemission, as in Figs. \ref{fig:1}{e},f. Data from the current-carrying device was collected while applying a constant current.  The energy and angular resolution were set to 90~meV and 0.2$^{\circ}$, respectively.  The spatial resolution was determined to be $(500 \pm 100)$~nm from a scan across the edge of the graphene flake (see illustration in Ref. \cite{Ulstrup:2019aa}). 
 The analyzer was aligned such that the cuts through the Dirac cone were approximately perpendicular to the $\Gamma-K$ direction. This avoids the suppression of one branch by matrix element effects  \cite{Shirley:1995aa,Mucha-Kruczynski:2008aa,Lizzit:2010aa,Gierz:2011ab}. The sample was carefully aligned  by acquiring an angle scan, such that the direction of measurement ($k_x=1.7$) was passing exactly through the Dirac point. 


\section{Appendix}

\subsection{2D fitting of spectral function}

Within 1~eV of the Fermi energy, the energy- and $k$-dependent photoemission intensity across the device contains only spectroscopic features from graphene and can therefore be well-described by a single Dirac cone. In order to determine the details of the dispersion and linewidth from the data, it proves advantageous to simultaneously fit the entire two-dimensional (2D) photoemission intensity to a model rather than to limit the analysis to fitting energy distribution curves or momentum distribution curves. Using this 2D fitting approach, we make best use of all the available information in the data and avoid spurious line broadenings close to band crossing points  \cite{Nechaev:2009aa}. Similar approaches have been employed in, e.g., Refs. \cite{Nechaev:2009aa,Bianchi:2010aa,Mazzola:2013aa,Andreatta:2019aa}.

In ARPES, the photoemission intensity is proportional to the hole spectral function  $ \mathcal{A}(E_b,\mathbf{k})$ multiplied by the Fermi-Dirac distribution. The spectral function is given in terms of the unrenormalized dispersion $\epsilon(\mathbf{k})$ and the complex self-energy $\Sigma=\Sigma' + i \Sigma''$. One usually assumes that $\Sigma$ depends on the energy $E_b$ but is independent of  $\mathbf{k}$.  $ \mathcal{A}(E_b,\mathbf{k})$    is given by 
 \begin{equation}
  \mathcal{A}(E_b,\mathbf{k})=\frac{\pi^{-1}|\Sigma''(E_b)|}{[ E_b-\epsilon(\mathbf{k})-\Sigma'(E_b)]^{2}+\Sigma''(E_b)^{2}}.
 \label{equ:p7}
 \end{equation}
Structure in the self-energy can describe subtle effects such as the band narrowing and kinks close to the Fermi energy due to electron-phonon coupling \cite{Hofmann:2009ab}. However, the electron-phonon coupling for undoped graphene is very weak \cite{Calandra:2007aa} and we thus  assume a simple model for the self-energy in which  $\Sigma'=0$ and $\Sigma''$ is an energy-independent constant. Such a model is expected to account well for the energy-independent electron-defect scattering. 

\begin{figure} [t!]
\includegraphics[width=0.5\textwidth]{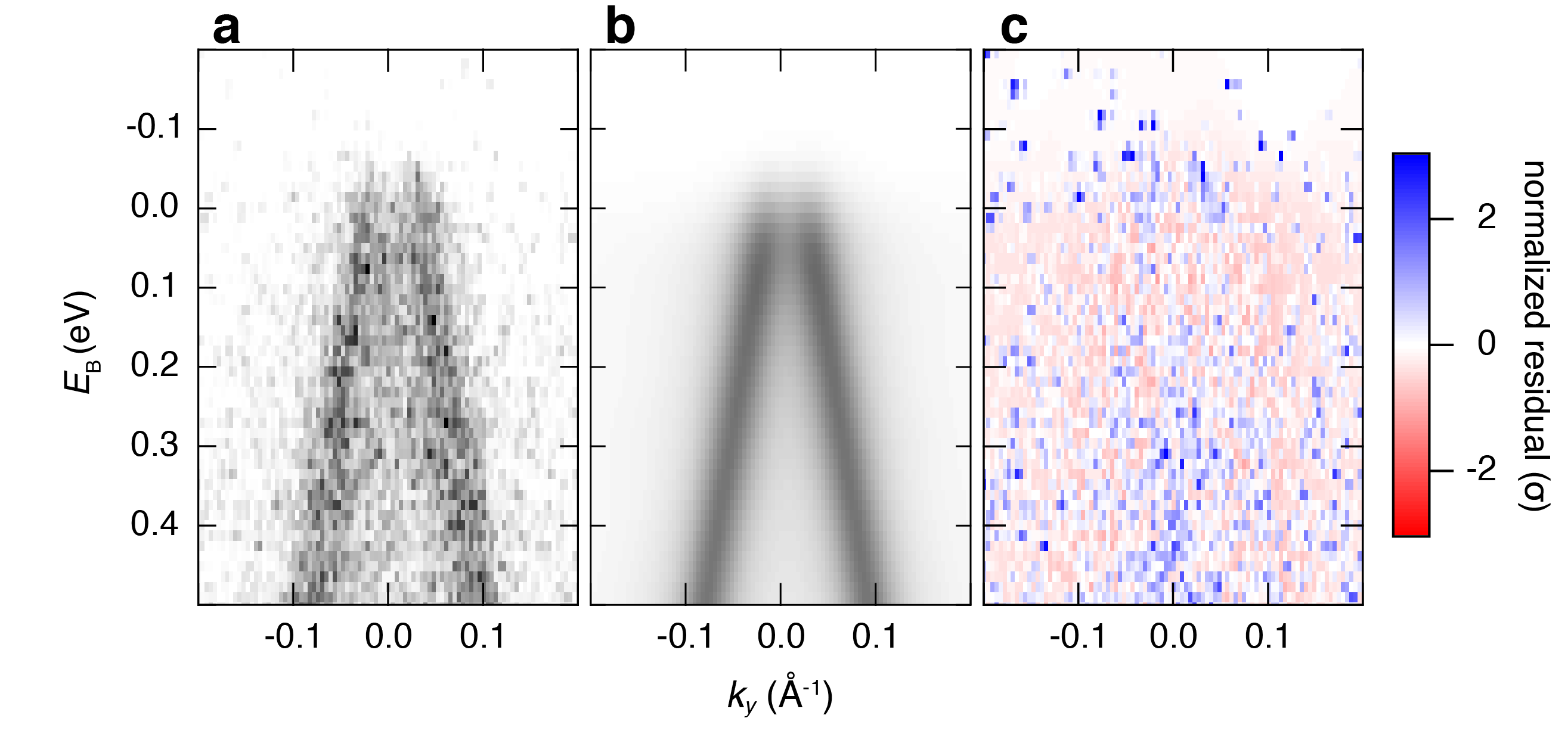}\\
\caption{\textbf{a} Spectrum from a single position on the device. \textbf{b} Fit as described in the text. \textbf{c} Difference between data and fit, divided by the square root of the intensity of the fit. Assuming Poissonian statistics for the noise, 95\% of the normalized residual pixels should have values between -2 and 2.}
 \label{fig:1s}
\end{figure}

We use a 2D cut through the spectral function in the  $E_b$-$k_y$ plane. The dispersion $\epsilon(k_y)$ is derived from a cut through a Dirac cone. A linear dispersion for the two resulting branches is assumed. The total range of the fit below the Fermi level is 500~meV. We assume that the dispersion is measured strictly along $k_y$ with $k_x=1.7$~\AA$^{-1}$, neglecting the slight curvature of the actually measured cut in  $\mathbf{k}$-space which is seen in Fig. \ref{fig:3}. The quality of the resulting fits justifies the use of this simple model (see Fig. \ref{fig:1s}). 

Photoemission matrix elements are not included in the formalism, but are important in ARPES from graphene due to 
 sub-lattice interference \cite{Shirley:1995aa,Mucha-Kruczynski:2008aa,Lizzit:2010aa,Gierz:2011ab}. For a momentum cut in the $\Gamma-K$ direction, this interference completely suppresses one branch of the Dirac cone. For a cut perpendicular to $\Gamma-K$, as we approximately have in this paper, the intensity of the two branches is the same. The combination of trigonal warping and matrix effects leads to the well-known horseshoe shape of the constant energy surfaces in ARPES from graphite and graphene, illustrated in Fig. \ref{fig:3} and in Fig. \ref{fig:3s}a. In order to account for any deviation of the scan direction from perpendicular to $\Gamma-K$, we treat the two Dirac cone branches  separately with respect to their photoemission intensity. This permits, in principle, information about the azimuthal rotation of the graphene flake from the intensity ratio between left and right branch to be extracted. 
 
 For the detailed analysis of local device properties shown in Figs. \ref{fig:3} and \ref{fig:4}, spectra were collected on a rectangular grid across the device with a point spacing of 250~nm in both directions. The spectra in Fig. \ref{fig:2} are generated by a sum of 9 such spectra in a 750$\times$750~nm$^2$ area. Fits to the model spectral function were performed for spectra taken on single grid points (see Fig. \ref{fig:1s}). 

Finally, the experimental energy- and $k$-broadening is taken into account by convoluting the product of spectral function and Fermi-Dirac distribution with the experimental resolution functions. For the analysis of the equilibrium properties, these are assumed to be Gaussians with a full width at half maximum of 90~meV and 0.0134~\AA$^{-1}$, respectively. A typical spectrum from a single position on the sample, along with the corresponding fit and residual, are shown in Fig. \ref{fig:1s}. The presence of the current leads to a degrading of the resolution because even a current of 0.5~mA through the device corresponds to a strong electric field of $\approx 5 \times 10^5$~V/m. The fit in the presence of the current is thus performed in two stages: We first set the energy resolution to 270~meV and fit all spectra. We then re-fit all spectra with an energy resolution corresponding to the expected energy broadening, given the electric field from the first fit and a UV spot size of 500~nm. 

\subsection{Discussion of different regions on the sample}

\begin{figure} [t!]
\includegraphics[width=0.5\textwidth]{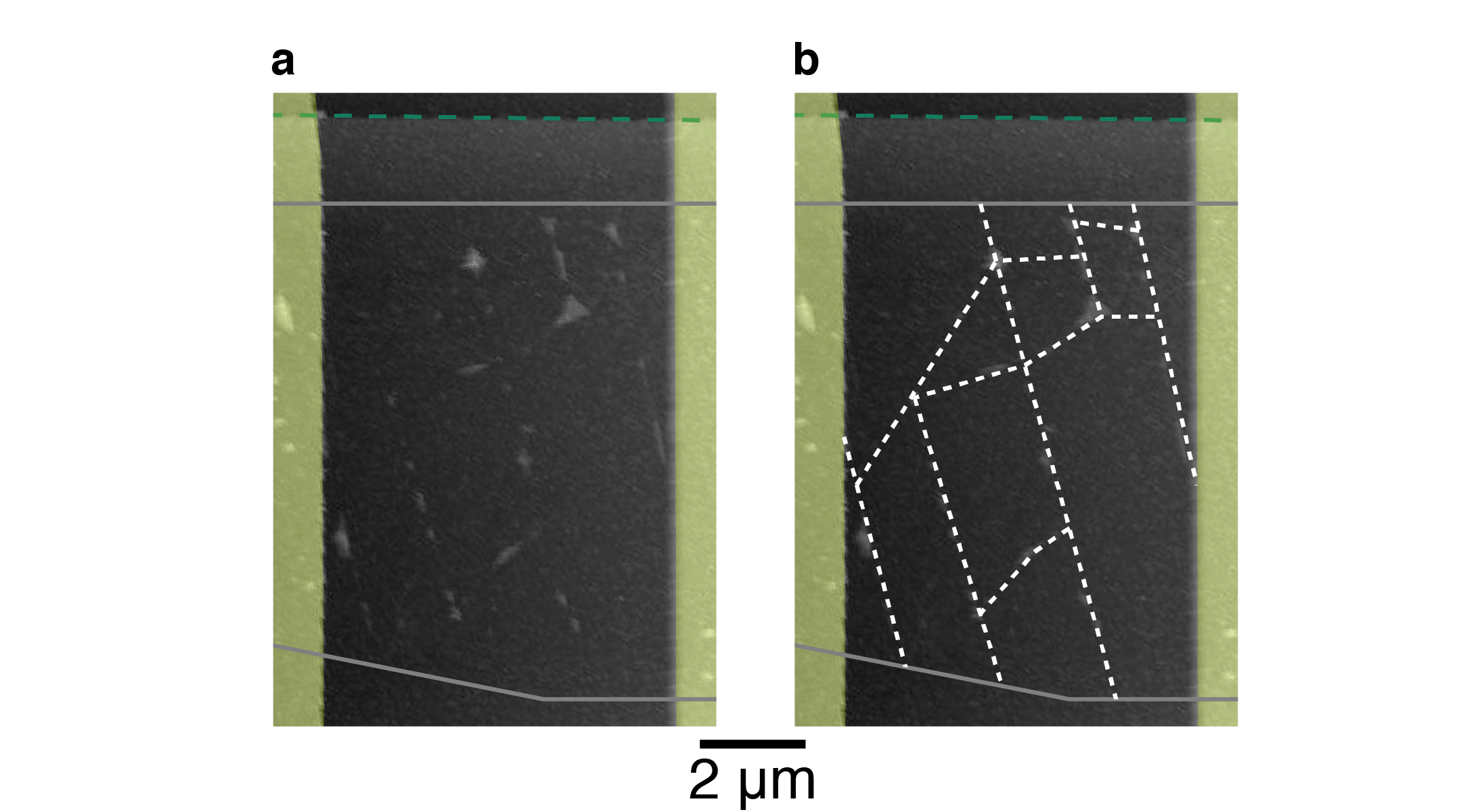}\\
\caption{\textbf{a} AFM image from Fig. \ref{fig:1}c without the dashed white lines, showing defect locations on the surface. \textbf{b} The same but with the white lines tracking the location of the defects. }
 \label{fig:2s}
\end{figure}

The dashed lines on the AFM image in \ref{fig:1}c indicate lines of protrusions, most likely caused by wrinkles in the graphene sheet, identified on the surface of the device. In order to see this more clearly, Fig. \ref{fig:2s} shows the AFM image with and without the dashed lines. 

\begin{figure} [h!]
\includegraphics[width=0.4\textwidth]{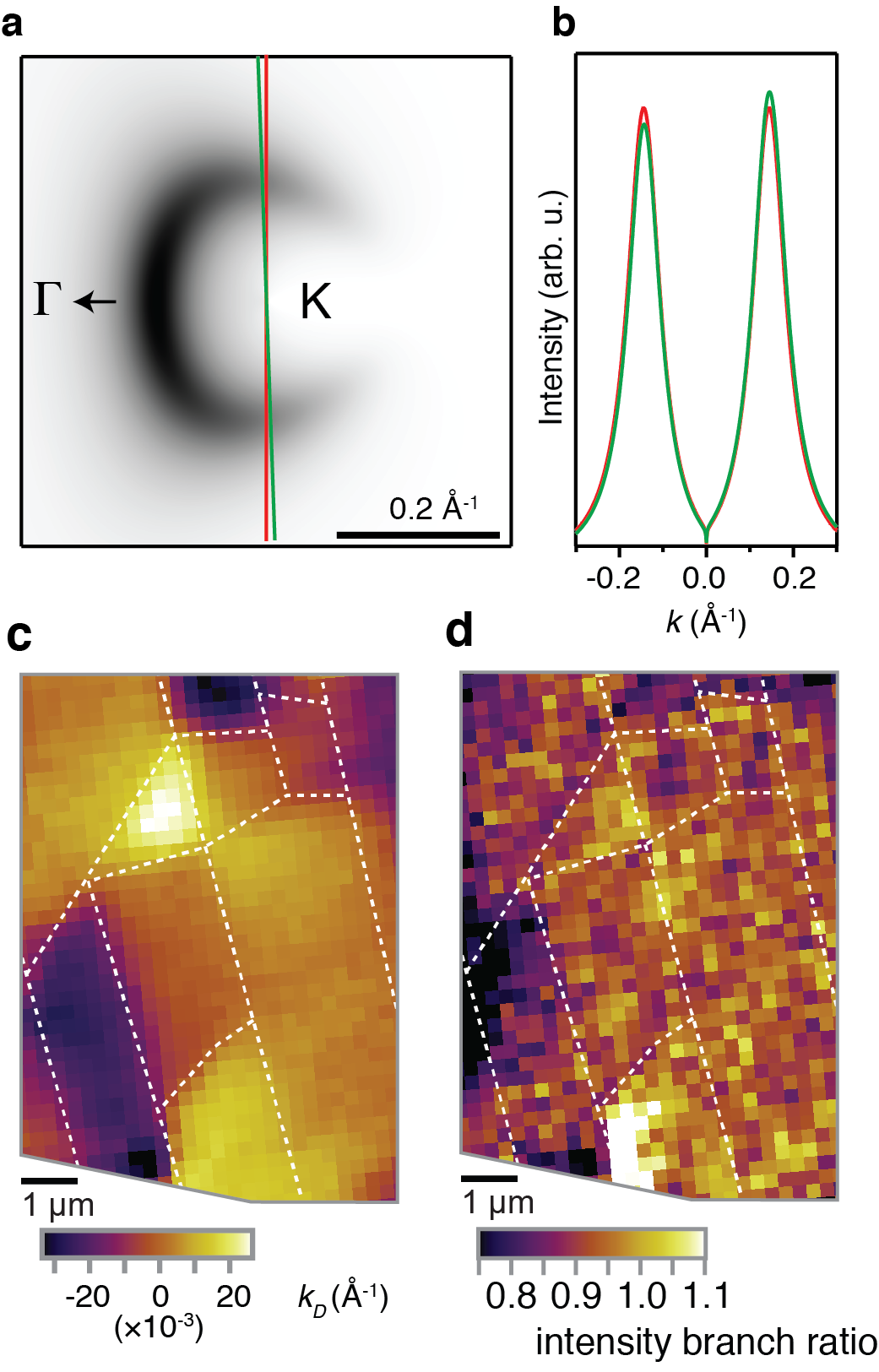}\\
\caption{\textbf{a} Simulated photoemission intensity distribution near the K-point of graphene for a constant energy cut 1 eV below the Dirac point. The  lines represent two directions enclosing an angle of 2$^{\circ}$. \textbf{b} Intensity distribution along the two lines in a. Note the small intensity difference between the two resulting lines in the case of the less symmetric cut.  \textbf{c} Map of $k_y$ shift from Fig. \ref{fig:3}b. \textbf{d} Map of the intensity ratio of the Dirac cone's left and right branch.}
 \label{fig:3s}
\end{figure}

Fig. \ref{fig:3}b shows the existence of different domains in the device, mainly distinguished by shifts in $k_y$. The domain structure in $k_y$ corresponds very well to the white dashed lines in Fig. \ref{fig:1} and in Fig. \ref{fig:2s}.
Naively, a variation of $k_y$ could suggest that the graphene flake contains domains with different surface normal directions. However, given the large size of the observed domains, this would result in considerable local height differences between the domains that cannot be reconciled with the atomically flat nature of the hBN substrate \cite{Dean:2010aa}. It is argued that the most likely origin of the $k_y$ shift is a slight rotation of domains against each other, typically by less than 1$^{\circ}$, caused by the formation of wrinkles in the graphene sheet. A domain rotation would not only be expected to lead to a $k_y$ shift of the Dirac cone in the window of observation, it should also result in a slight left/right asymmetry in the intensity of the two observed branches. This is illustrated in Fig. \ref{fig:3s}. Fig. \ref{fig:3s}{a} shows the expected photoemission intensity at constant energy 1 eV below the Dirac point. The calculation includes trigonal warping and sub-lattice interference effects as in Ref. \cite{Lizzit:2010aa}, resulting in the characteristic horseshoe intensity shape discussed in connection with the spectral function, above. Fig. \ref{fig:3s}{b}  shows two cuts through this intensity distribution taken at slightly different angles, marked by the lines in Fig. \ref{fig:3s}a. The intensity distribution for the symmetric cut, exactly perpendicular to the $\Gamma$-K direction (red line), is perfectly symmetric but the distribution for a cut rotated by an angle of 2$^{\circ}$ (green line) shows a slight intensity asymmetry between the branches (the peak intensity difference for this angle is 6\%). 

A careful inspection of the data does indeed confirm the existence of an intensity change consistent with the interpretation of the $k_y$ shift caused by domain rotations. Fig. \ref{fig:3s}{c} shows the $k_y$ shift as in Fig. \ref{fig:3}b and  Fig. \ref{fig:3s}{d} gives the corresponding ratio between left and right branches of the Dirac cone. Despite the very small size of the effect compared to the noise level in the data, a clear correlation exists between these images. The size of the shift in $k_y$ also can be converted into the corresponding rotation of the domains. For the entire spread of $k_y$ values in the figure, approximately a range of $\Delta k_y=0.06$~\AA$^{-1}$, this corresponds to a total angular range of  $2^{\circ}$, since $\Delta k_y = (\Gamma-K) \sin \phi$, where $(\Gamma-K)$ is the distance between $\Gamma$ and $K$ in graphene (1.7~\AA$^{-1}$) and $\phi$ is the azimuthal rotation angle between the flakes. 

\subsection{Calculation of conductivity maps}

The calculation of the conductivity maps from the potential landscape across the sample is an ill-posed problem that has been addressed in detail in connection with scanning probe potentiometry. Here we apply the approach proposed by Zhang et al. in Ref. \cite{Zhang:2016ag} with some minor modifications. Given the local potential $\phi$, the local conductivity $\sigma$ has to fulfil both the modified  Poisson equation 
\begin{equation} 
\nabla \cdot (\sigma \cdot \nabla \phi) = 0,
\end{equation}
and, in the absence of a magnetic field and for diffusive transport, 
\begin{equation}
 \nabla \times (\sigma \cdot \nabla \phi) = 0.
 \end{equation}
These equations can be combined, resulting in an underdetermined linear system of equations that is then approximately solved by the search for an optimum solution $\sigma'$ which we here assume to be locally isotropic i.e., the conductivity at each point on the graphene flake is independent of the current direction through it. It has been shown that the best solution for    $\sigma'$ cannot be found efficiently by a steepest descent method \cite{Zhang:2016ag}. Therefore, and in order to avoid the risk of finding a $\sigma'$ not corresponding to the global optimum, we use a combination of simulated annealing and steepest descent strategies in the optimization process \cite{Press:2007aa}. The procedure can only give relative conductivity values but these can be normalized using the known total conductance of the device. 

\subsection{Data for very high current densities}

\begin{figure} [h!]
\includegraphics[width=0.5\textwidth]{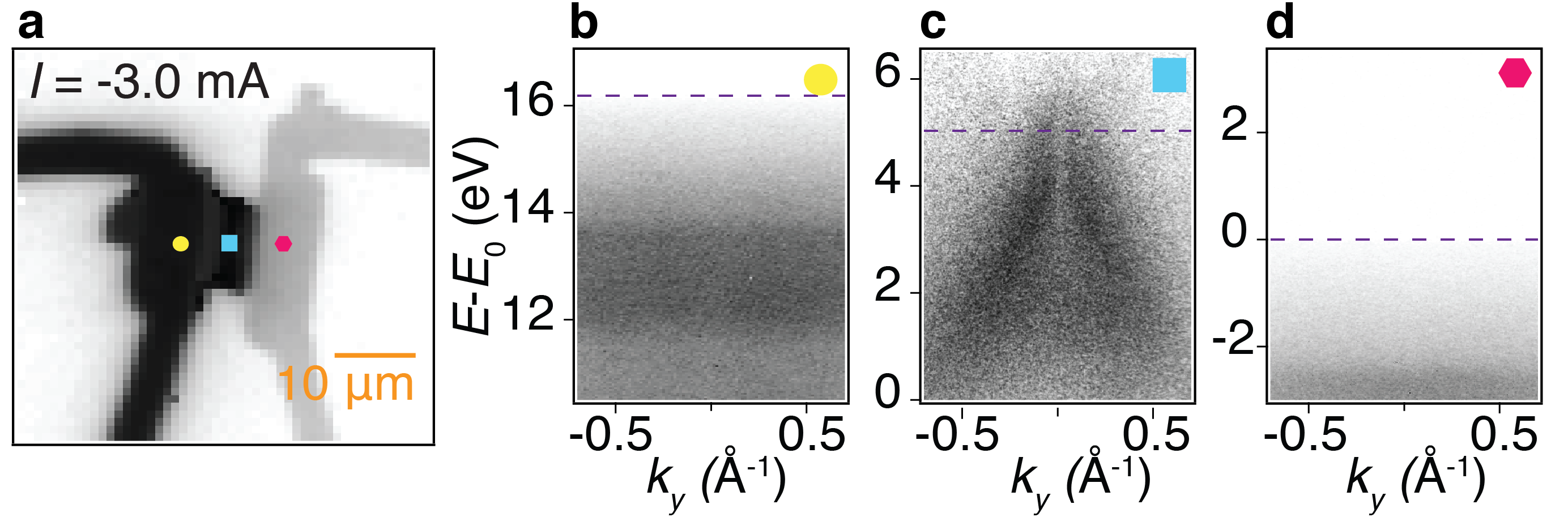}\\
\caption{Data collected close to the breakdown current density of graphene (sample current of 3~mA).  \textbf{a} Integrated photoemission intensity from the device. \textbf{b}-\textbf{d} Spectra taken at the positions indicated by coloured markers in a. The energy zero $E_0$ is chosen to be the Fermi energy on the right hand electrode.}
 \label{fig:4s}
\end{figure}

It is possible to apply an electric field to the sample that is sufficiently high to considerably degrade the energy resolution in the experiment, even when using a very small light spot as we do here. This is due to the ability of graphene to withstand the resulting very high current density. Fig. \ref{fig:4s}a  shows the integrated photoemission intensity from the device for a current of 3~mA, corresponding to a 2D current density of $j=$1.9~Acm$^{-1}$ or, assuming a graphene thickness of 0.35~nm, to a 3D current density  of 5.2$\times$10$^{7}$~Acm$^{-2}$, close to the breakdown current density of graphene   \cite{Moser:2007aa,Murali:2009aa,Leitherer:2019aa}. In this case, the voltage difference between the Au electrodes is so large that  electrons emitted from the right hand side electrode are mostly outside the window of detectable energies,  and  this electrode thus appears much lighter than the one on the left hand side. Fig. \ref{fig:4s}b-d show spectra from the left hand electrode, the centre and the right hand electrode with the Fermi levels indicated (the position for the spectra are marked in Fig. \ref{fig:4s}a). The total potential difference over the device is 16.2~V. This leads to a strong energy broadening for the graphene spectra within the 500~nm light spot, as seen in Fig. \ref{fig:4s}c. Note that the data for this current density were taken with a high pass energy of the electron analyzer ({100~eV}) to capture the broad energy range of interest. The measured energy resolution from the Au contacts with an applied field is 90~meV. This is insufficient to explain the significantly larger energy broadening of the graphene spectrum that is caused by the voltage drop within the light spot area. 

\section{acknowledgement}
We thank Diamond Light Source for access to Beamline I05 (Proposal No. SI20218 and SI20078) that contributed to the results presented here. S.U. acknowledges financial support from VILLUM FONDEN under the Young Investigator Program (grant no. 15375).  This work was supported by VILLUM FONDEN via the Centre of Excellence for Dirac Materials (Grant No. 11744). J. A. M. acknowledges financial support from the Danish Council for Independent Research, Natural Sciences under the Sapere Aude program (Grant No. DFF-6108-00409). Growth of hBN crystals was supported by the
Elemental Strategy Initiative conducted by the MEXT, Japan and the
CREST(JPMJCR15F3), JST.
R. M. and J. K. acknowledges the financial support from U.S. Department of Energy, Office of Science, Office of Basic Energy Sciences, under Award Number DE-SC0020323 and partial support from the Center for Emergent materials: an NSF MRSEC under award number DMR-1420451.
%
%


\end{document}